\documentstyle[aps,prd]{revtex}

\begin{document}
\title{Mixing angle and Glashow's Algebra}
\author{A. L. Barbosa}
\address{Laboratoire de Gravitation et Cosmologie \\
Relativistes \\
Universit{\'e} Pierre et Marie Curie, CNRS/ESA 7065 \\
4, Place Jussieu \ Cedex 05 \\
75252 Paris, France}
\date{\today}

\maketitle

\begin{abstract}

Considering transformations in the basis of fundamental fields on a 
principal fiber bundle, without modification in the space-time sector, we 
construct an algebra {\itshape GA}, which we call Glashow algebra. 
The structure constants of this algebra depend on a mixing angle. 
The Lagrangian of the gauge theory of electroweak interactions without masses is obtained using a 
representation of {\itshape GA} which is the transformed of the 
adjoint representation of $SU(2)\otimes U(1)$, and does not coincide with the 
adjoint representation of {\itshape GA}. The mixing angle is automatically present in 
the theory if {\itshape GA} is used.

\end{abstract}

\section{Introduction}

All the fundamental interactions, with the exception of gravitation, are
described by gauge theories. Such is the case for the well established QED,
for the Standard model and for QCD. From the geometrical point of view such
theories correspond, in general terms, to the construction of a principal
fiber bundle with the corresponding gauge group as the structure group and
space-time as the base manifold (Aldrovandi and Pereira, 1995). The choice of a connection,
which establishes the unicity of decomposition of any vector field on the
space tangent to the bundle into a vertical and a horizontal components,
corresponds to the introduction of a gauge potential. The well-known
transformation of connections by the action of the group (adjoint
representation), when pulled back to space-time, gives rise to the usual
transformation of gauge fields. In each associated fiber bundle the covariant
derivative defined in the principal fiber bundle will acquire a form
corresponding to the representation of the fields to which it is being
applied. That form is exactly what comes up when the interactions are
introduced through the minimal coupling prescription.

In consequence, a gauge theory can be locally described on the bundle by the
following commutation relations (Cho, 1975)
\begin{equation}%
\begin{array}
[c]{c}%
\left[  D_{\mu},D_{\nu}\right]  =-F_{\hspace{0.1cm}\mu\nu}^{a}X_{a},\\
\\
\left[  D_{\mu},X_{a}\right]  =0,\\
\\
\left[  X_{a},X_{b}\right]  =f_{\hspace{0.1cm}ab}^{c}X_{c},
\end{array}
\label{a53}%
\end{equation}
where $X_{a}$'s are the fundamental vector fields on bundle, which represent
the generators of the algebra of the structure group, and $D_{\mu}$ is the
covariant derivative given by
\begin{equation}
D_{\mu}=\partial_{\mu}-gA_{\hspace{0.1cm}\mu}^{a}X_{a}\label{a54}%
\end{equation}
with $A_{\hspace{0.1cm}\mu}^{a}$ being the gauge field, which is a 
connection, and g is a coupling constant.
$F_{\hspace{0.1cm}\mu\nu}^{a}$ is the field strength of the gauge field,
\begin{equation}
F_{\hspace{0.1cm}\mu\nu}^{a}=g\left[ \partial_{\mu}A_{\hspace{0.1cm}\nu}^{a}%
-\partial_{\nu}A_{\hspace{0.1cm}\mu}^{a}+gf_{\hspace{0.1cm}bc}^{a}%
A_{\hspace{0.1cm}\mu}^{b}A_{\hspace{0.1cm}\nu}^{c}\label{a55} \right]%
\end{equation}

Three of the four interaction of Nature are described by gauge theories. For
electromagnetism, the structure group is $U(1),$ an abelian group and the
commutation relations and field strength are respectively
\begin{equation}%
\begin{array}
[c]{c}%
\left[  D_{\mu},D_{\nu}\right]  =-F_{\mu\nu},\\
\\
\left[  D_{\mu},X_{0}\right]  =0,\\
\\
\left[  X_{0},X_{0}\right]  =0,
\end{array}
\end{equation}
and
\begin{equation}
F_{\mu\nu}=\partial_{\mu}A_{\nu}-\partial_{\nu}A_{\mu},
\end{equation}
where $X_{0}$ is the generator of $U(1).$

For QCD (Cheng and Li, 1984) the structure group is $SU(3)$ which has eight
generators and, in the Gell-Mann basis, the following structure constants
\begin{equation}%
\begin{array}
[c]{c}%
f_{\hspace{0.1cm}12}^{3}=1,f_{\hspace{0.1cm}41}^{7}=\frac{1}{2},f_{\hspace
{0.1cm}15}^{6}=-\frac{1}{2},f_{\hspace{0.1cm}24}^{6}=\frac{1}{2}%
,f_{\hspace{0.1cm}25}^{7}=\frac{1}{2},\\
\\
f_{\hspace{0.1cm}34}^{5}=\frac{1}{2},f_{\hspace{0.1cm}36}^{7}=-\frac{1}%
{2},f_{\hspace{0.1cm}45}^{8}=\frac{\sqrt{3}}{2},f_{\hspace{0.1cm}67}^{8}%
=\frac{\sqrt{3}}{2}.
\end{array}
\label{a56}%
\end{equation}
The commutation relations and field strength are directly obtained using
$\left(  \ref{a56}\right)  $ in $\left(  \ref{a53}\right)  $ and $\left(
\ref{a55}\right)  $.

In the case of electroweak interactions, the gauge group is not $SU(2)\otimes
U(1)$ at all, because to obtain the couplings a mixing angle must be
introduced. We cannot write for this theory the structure constants as we
wrote above for QCD and QED.

In the present paper we examine the introduction of a mixing angle in a
non-abelian gauge theory through a modification of the algebra which makes it
possible to write the structure constants in a way analogous to that of QCD
and QED.This leads to a new algebra which we call \textit{Glashow algebra
(GA)}, giving a geometrical interpretation for the introduction of the mixing
angle in electroweak theory. This means that we obtain the Lie algebra
corresponding to $SU(2)\otimes U(1)\mid_{mixed}$. The usual Lagrangian of
gauge theories is obtained by taking the trace . We do obtain the Lagrangian
of the Glashow model (electroweak interactions with no massive 
bosons) (Mandl and Shaw 1984) in that way.
Notice that it is not at all evident that this can be done, as the algebra
\textit{GA }is non-semisimple.

In section 2 we present the construction of the representations of the direct
product $SU(2)\otimes U(1)$ which is needed to obtain some representations of
the Glashow algebra. In section 3, we construct the Glashow algebra and in
section 4 we obtain three representations for it. Section 5 contains the
calculus of traces and the construction of the Lagrangian for the Glashow
model. Section 6 is reserved for conclusion and final remarks.

\section{Construction of the representations of the direct product
$SU(2)\otimes U(1)$}

Let $\left\{  X_{a}\right\}$ be the three generators of $SU(2)$ and $X_{0}$
that of $U(1)$. We now proceed to construct the representations associated to
the direct product of $SU(2)$ and $U(1).$

Consider the commutation relations of both algebras separately:%

\begin{equation}
[ X_{a},X_{b}]=\epsilon_{\hspace{0.1cm}ab}^{c}X_{c},\label{a17}%
\end{equation}%
\begin{equation}
[ X_{a},X_{b}]=0,\textnormal{ for }a\ \textnormal{or }b=0.\label{a18}%
\end{equation}

For the fundamental representation we use the condition
\begin{equation}
tr[\left(  T_{a}\right)  ^{2}]=-\frac{1}{2}.\label{a19}%
\end{equation}
It follows that the fundamental representation is composed by $3\times3$
matrices whose squares have trace equal to $-\frac{1}{2}.$ Condition $\left(
\ref{a19}\right)  $ characterizes the the fundamental representation of $SU(2)
$ alone. We shall keep the same condition to get a representation which
extends that representation. The generators are%

\begin{equation}
T_{1}=\left[
\begin{array}
[c]{ccc}%
0 & -\frac{i}{2} & 0\\
-\frac{i}{2} & 0 & 0\\
0 & 0 & 0
\end{array}
\right], \label{a20}%
\end{equation}%
\begin{equation}
T_{2}=\left[
\begin{array}
[c]{ccc}%
0 & -\frac{1}{2} & 0\\
\frac{1}{2} & 0 & 0\\
0 & 0 & 0
\end{array}
\right], \label{a21}%
\end{equation}%
\begin{equation}
T_{3}=\left[
\begin{array}
[c]{ccc}%
-\frac{i}{2} & 0 & 0\\
0 & \frac{i}{2} & 0\\
0 & 0 & 0
\end{array}
\right], \label{a22}%
\end{equation}%
\begin{equation}
T_{0}=\left[
\begin{array}
[c]{ccc}%
0 & 0 & 0\\
0 & 0 & 0\\
0 & 0 & \frac{i}{\sqrt{2}}%
\end{array}
\right] .\label{a23}%
\end{equation}

All these matrices satisfy $\left(  \ref{a17}\right)  $, $\left(
\ref{a18}\right)  $, $\left(  \ref{a19}\right)  $ and can be taken to generate
the fundamental representation of the direct product $SU(2)\otimes U(1).$

For the adjoint representation we have $4\times4$\ matrices, whose squares
have traces
\begin{equation}
tr[\left(  X_{a}\right)  ^{2}]=-2,\label{a24}%
\end{equation}
a condition which is also valid for the adjoint representation of $SU(2).$

We obtain the matrices
\begin{equation}
X_{1}=\left[
\begin{array}
[c]{cccc}%
0 & 0 & 0 & 0\\
0 & 0 & -1 & 0\\
0 & 1 & 0 & 0\\
0 & 0 & 0 & 0
\end{array}
\right],  \label{a25}%
\end{equation}%
\begin{equation}
X_{2}=\left[
\begin{array}
[c]{cccc}%
0 & 0 & 1 & 0\\
0 & 0 & 0 & 0\\
-1 & 0 & 0 & 0\\
0 & 0 & 0 & 0
\end{array}
\right],  \label{a26}%
\end{equation}%
\begin{equation}
X_{3}=\left[
\begin{array}
[c]{cccc}%
0 & -1 & 0 & 0\\
1 & 0 & 0 & 0\\
0 & 0 & 0 & 0\\
0 & 0 & 0 & 0
\end{array}
\right],  \label{a27}%
\end{equation}%
\begin{equation}
X_{0}=\left[
\begin{array}
[c]{cccc}%
0 & 0 & 0 & 0\\
0 & 0 & 0 & 0\\
0 & 0 & 0 & 0\\
0 & 0 & 0 & i\sqrt{2}%
\end{array}
\right], \label{a28}%
\end{equation}
satisfying $\left(  \ref{a17}\right)  $, $\left(  \ref{a18}\right)  $ and
$\ref{a24}.$ They generate the adjoint representation of the direct product.

\section{The Glashow algebra}

Suppose we have a gauge theory whose commutation relations are given by
$\left(  \ref{a53}\right)  $ and whose gauge potentials (or connections)\ and
field strength are respectively
\begin{equation}
A_{\mu}=A_{\hspace{0.1cm}\mu}^{a}X_{a}\label{a61}%
\end{equation}
and $\left(  \ref{a55}\right)  $. On the fiber bundle, they transform as
\begin{equation}
X_{a}(A_{\hspace{0.1cm}\mu}^{b})=f_{\hspace{0.1cm}ac}^{b}A_{\hspace{0.1cm}\mu
}^{c},\label{a57}%
\end{equation}%
\begin{equation}
X_{a}(F_{\hspace{0.1cm}\mu\nu}^{b})=f_{\hspace{0.1cm}ac}^{b}F_{\hspace
{0.1cm}\mu\nu}^{c}.\label{a58}%
\end{equation}

For the direct product $SU(2)\otimes U(1),$ we observe that there is no
charged fields in the theory. Interaction terms involve $f_{\hspace{0.1cm}%
ac}^{b}.$ In consequence, we see from $\left(  \ref{a18}\right)  $ that there
exists no interaction terms between the abelian and nonabelian sector. From
the experimental data on electroweak interactions we know that there are two
charged bosons $W^{+}$ and $W^{-}$ and that there is a mixture between the
abelian and nonabelian sector, which gives an essential contribution to the
cross section of the scattering (Mandl and Shaw 1984), (Greiner and 
M\"uller, 1996)
\begin{equation}
e^{+}+e^{-}\longrightarrow W^{+}+W^{-}.
\end{equation}
In the usual gauge theory for electroweak interaction this problem is solved
by introducing a mixing angle directly in the Lagrangian. The physical fields
appear as mixtures of the original gauge potentials.

Our aim is to interpret the introduction of the mixing angle from the
algebraic point of view. We shall obtain an usual gauge theory considering not
the direct product algebra but another algebra \textit{GA}\ which we will
construct. In order to achieve this aim we should answer the following question:

\textit{What is the set of commutation relations which corresponds to the
}$SU(2)\otimes U(1)\mid_{mixed}$?

We call $\left\{  X_{a}\right\}  $ the basis of fields corresponding to the
gauge fields $A_{\mu}=A_{\hspace{0.1cm}\mu}^{a}X_{a},$ which satisfy the
commutation relations
\begin{equation}
[ X_{a},X_{b}]=f_{\hspace{0.1cm}ab}^{c}X_{c}.\label{a29}%
\end{equation}

We associate a new basis of fields $\left\{  X_{a}^{\prime}\right\}  $ to the
physical potentials $A_{\hspace{0.1cm}\mu}^{\prime a}$ which include two
charged and two neutral fields$:$
\begin{equation}
A_{\mu}^{\prime}=A_{\hspace{0.1cm}\mu}^{\prime a}X_{a}^{\prime}.\label{a30}%
\end{equation}

The modification has been made only in the algebraic sector, consequently we
expect to have no change in space-time
\begin{equation}
A_{\mu}^{\prime}=A_{\mu}.\label{a31}%
\end{equation}

The neutral fields are $A_{\hspace{0.1cm}\mu}^{\prime0}$ and $A_{\hspace
{0.1cm}\mu}^{\prime3},$ while the charged fields are given by
\begin{equation}
A_{\hspace{0.1cm}\mu}^{\prime1}=\frac{1}{\sqrt{2}}\left(  A_{\hspace{0.1cm}%
\mu}^{1}-iA_{\hspace{0.1cm}\mu}^{2}\right)  \textnormal{ or }A_{\hspace{0.1cm}\mu
}^{1}=\frac{1}{\sqrt{2}}\left(  A_{\hspace{0.1cm}\mu}^{\prime1}+A_{\hspace
{0.1cm}\mu}^{\prime2}\right)  ,\label{a32}%
\end{equation}%
\begin{equation}
A_{\hspace{0.1cm}\mu}^{\prime2}=\frac{1}{\sqrt{2}}\left(  A_{\hspace{0.1cm}%
\mu}^{1}+iA_{\hspace{0.1cm}\mu}^{2}\right)  \textnormal{ or }A_{\hspace{0.1cm}\mu
}^{2}=\frac{i}{\sqrt{2}}\left(  A_{\hspace{0.1cm}\mu}^{\prime1}-A_{\hspace
{0.1cm}\mu}^{\prime2}\right)  .\label{a33}%
\end{equation}

To consider the charged fields in the direct product $SU(2)\otimes U(1)$ is
not enough to produce the correct couplings between the fields. We observe in
the Lagrangian or in the equation of motion that the absence of interaction
between $A_{\hspace{0.1cm}\mu}^{\prime0}$ and the other components is due to
the values of the structure constants.We must therefore, modify the algebra to
obtain the couplings. For that we begin by making the following
transformation
\begin{equation}%
\begin{array}
[c]{c}%
A_{\hspace{0.1cm}\mu}^{3}=\alpha A_{\hspace{0.1cm}\mu}^{\prime3}+\beta
A_{\hspace{0.1cm}\mu}^{\prime0},\\
\\
A_{\hspace{0.1cm}\mu}^{0}=\gamma A_{\hspace{0.1cm}\mu}^{\prime3}+\delta
A_{\hspace{0.1cm}\mu}^{\prime0}.
\end{array}
\label{a34}%
\end{equation}

In order to determine $\alpha$,$\beta$,$\gamma$ and $\delta$ we impose on
$\left(  \ref{a34}\right)  $ the following conditions:

\begin{enumerate}
\item  Preservation of the quadratic terms: $\left(  A_{\hspace{0.1cm}\mu}%
^{0}\right)  ^{2}+\left(  A_{\hspace{0.1cm}\mu}^{3}\right)  ^{2}=\left(
A_{\hspace{0.1cm}\mu}^{\prime0}\right)  ^{2}+\left(  A_{\hspace{0.1cm}\mu
}^{\prime3}\right)  ^{2}$. This condition is imposed as if there were 
mass terms in the Lagrangian, which we want to preserve.

\item  Transformation continuously connected to the identity.
\end{enumerate}

This leads to the conditions:
\begin{equation}
\alpha\beta+\gamma\delta=0,\label{a35}%
\end{equation}%
\begin{equation}
\alpha^{2}+\gamma^{2}=1,\label{a36}%
\end{equation}%
\begin{equation}
\beta^{2}+\delta^{2}=1,\label{a37}%
\end{equation}%
\begin{equation}
\alpha\delta-\gamma\beta=1,\label{a38}%
\end{equation}
from which result the following cases:

\begin{enumerate}
\item $\delta=\alpha,\beta=-\gamma\Rightarrow\alpha^{2}+\gamma^{2}=1. $ For
which we choose the parametrization

$\alpha=\delta=\cos\theta$

$\beta=-\gamma=\sin\theta$

\item $\delta=\alpha$, $\beta=\gamma\Rightarrow\alpha^{2}=1,$ $\gamma=0,$
$\beta=0.$

\item $\delta=-\alpha$, $\beta=-\gamma\Rightarrow\gamma^{2}=1,$ $\alpha=0,$
$\delta=0.$
\end{enumerate}

We write (\ref{a34}) for the three cases above

\begin{itemize}
\item Case 1 
\begin{eqnarray}%
A^3{}_\mu=\cos\theta A_{\hspace{0.1cm}\mu}^{\prime3}%
+\sin\theta A_{\hspace{0.1cm}\mu}^{\prime0}, \nonumber \\
{}\label{a40} \\
A_{\hspace{0.1cm}\mu}^{0}=-\sin\theta A_{\hspace{0.1cm}\mu}^{\prime3}%
+\cos\theta A_{\hspace{0.1cm}\mu}^{\prime0} \nonumber.
\end{eqnarray}

\item Case 2
\begin{equation}%
\begin{array}
[c]{c}%
A_{\hspace{0.1cm}\mu}^{3}= \pm A_{\hspace{0.1cm}\mu}^{\prime3},\\
\\
A_{\hspace{0.1cm}\mu}^{0}= \pm A_{\hspace{0.1cm}\mu}^{\prime0}.
\end{array}
\label{a39}%
\end{equation}

\item Case 3
\begin{equation}%
\begin{array}
[c]{c}%
A_{\hspace{0.1cm}\mu}^{3}= \mp A_{\hspace{0.1cm}\mu}^{\prime0},\\
\\
A_{\hspace{0.1cm}\mu}^{0}= \pm A_{\hspace{0.1cm}\mu}^{\prime3}.
\end{array}
\label{a41}%
\end{equation}
\end{itemize}

Expressions $\left(  \ref{a40}\right)  $ correspond to the usual expressions
of mixing gauge fields in Weinberg-Salam theory and we will use it to
construct the new algebra.

Considering $\left(  \ref{a61}\right)  $ and $\left(  \ref{a30}\right)  ,$ we
write $\left(  \ref{a31}\right)  $ explicitly
\begin{equation}%
\begin{array}
[c]{c}%
A_{\hspace{0.1cm}\mu}^{\prime1}X_{1}^{\prime}+A_{\hspace{0.1cm}\mu}^{\prime
2}X_{2}^{\prime}+A_{\hspace{0.1cm}\mu}^{\prime3}X_{3}^{\prime}+A_{\hspace
{0.1cm}\mu}^{\prime0}X_{0}^{\prime}=\\
\\
A_{\hspace{0.1cm}\mu}^{1}X_{1}+A_{\hspace{0.1cm}\mu}^{2}X_{2}+A_{\hspace
{0.1cm}\mu}^{3}X_{3}+A_{\hspace{0.1cm}\mu}^{0}X_{0}.
\end{array}
\label{a42}%
\end{equation}
Using $\left(  \ref{a32}\right)  ,$ $\left(  \ref{a33}\right)  $ and $\left(
\ref{a40}\right)  $ and equating the coefficients of each component of
$A_{\hspace{0.1cm}\mu}^{\prime a},$ we obtain the new generators%
\begin{equation}
X_{1}^{\prime}=\frac{1}{\sqrt{2}}(X_{1}+iX_{2}),\label{a43}%
\end{equation}%
\begin{equation}
X_{2}^{\prime}=\frac{1}{\sqrt{2}}(X_{1}-iX_{2}),\label{a44}%
\end{equation}%
\begin{equation}
X_{3}^{\prime}=\cos\theta X_{3}-\sin\theta X_{0},\label{a45}%
\end{equation}%
\begin{equation}
X_{0}^{\prime}=\sin\theta X_{3}+\cos\theta X_{0}.\label{a46}%
\end{equation}

The new algebra is characterized by the commutation relations of the fields
$\left\{  X_{a}^{\prime}\right\}  $:
\begin{equation}%
\begin{array}
[c]{c}%
\left[  X_{1}^{\prime},X_{2}^{\prime}\right]  =-i\left(  \sin\theta
X_{0}^{\prime}+\cos\theta X_{3}^{\prime}\right)  ,\\
\\
\left[  X_{1}^{\prime},X_{3}^{\prime}\right]  =i\cos\theta X_{1}^{\prime},\\
\\
\left[  X_{1}^{\prime},X_{0}^{\prime}\right]  =i\sin\theta X_{1}^{\prime},\\
\\
\left[  X_{2}^{\prime},X_{3}^{\prime}\right]  =-i\cos\theta X_{2}^{\prime},\\
\\
\left[  X_{2}^{\prime},X_{0}^{\prime}\right]  =-i\sin\theta X_{2}^{\prime},\\
\\
\left[  X_{0}^{\prime},X_{3}^{\prime}\right]  =0.
\end{array}
\label{a47}%
\end{equation}

We observe that there is now a mixing between the generators of the two sectors.

Cases $2$ and $3$ above are particular cases of $1$. To case $2$ correspond
the angles $\theta=0$ and $\theta=\pi$, giving rise to the algebra
\begin{equation}%
\begin{array}
[c]{c}%
\left[  X_{1}^{\prime},X_{2}^{\prime}\right]  =\mp iX_{3}^{\prime},\\
\\
\left[  X_{1}^{\prime},X_{3}^{\prime}\right]  =\pm iX_{1}^{\prime},\\
\\
\left[  X_{1}^{\prime},X_{0}^{\prime}\right]  =0,\\
\\
\left[  X_{2}^{\prime},X_{3}^{\prime}\right]  =\mp iX_{2}^{\prime},\\
\\
\left[  X_{2}^{\prime},X_{0}^{\prime}\right]  =0,\\
\\
\left[  X_{0}^{\prime},X_{3}^{\prime}\right]  =0.
\end{array}
\end{equation}

For case $3$ we have $\theta=\frac{\pi}{2}$ and $\theta=\frac{3\pi}{2}$ and
the algebra is
\begin{equation}%
\begin{array}
[c]{c}%
\left[  X_{1}^{\prime},X_{2}^{\prime}\right]  =\mp iX_{0,}^{\prime}\\
\\
\left[  X_{1}^{\prime},X_{3}^{\prime}\right]  =0,\\
\\
\left[  X_{1}^{\prime},X_{0}^{\prime}\right]  =\pm iX_{1}^{\prime},\\
\\
\left[  X_{2}^{\prime},X_{3}^{\prime}\right]  =0,\\
\\
\left[  X_{2}^{\prime},X_{0}^{\prime}\right]  =\mp iX_{2}^{\prime},\\
\\
\left[  X_{0}^{\prime},X_{3}^{\prime}\right]  =0.
\end{array}
\end{equation}

These three sets of commutation relations satisfy Jacobi identities and
therefore constitute Lie algebras. As noticed, the last two algebras above are
particular cases of the most general one given by $\left(  \ref{a47}\right)
,$ which we call \textit{Glashow algebra }$\left(  \mathit{GA}\right)  $.

Since we do not have masses in the theory, all values of angles are admissible
and we could be tempted to construct a theory for each one of these algebras.
But with mass generation and considering the usual masses relations of the
Weinberg-Salam model, in wich $sin\theta$ appears in the denominator, algebras
$2$ and $3$ are actually excluded since they would imply infinite masses for
the gauge bosons.

The structure constants of the new algebra $\left(  \ref{a47}\right)  $ are
\begin{equation}%
\begin{array}
[c]{c}%
f_{\hspace{0.1cm}12}^{\prime0}=-i\sin\theta,\\
\\
f_{\hspace{0.1cm}12}^{\prime3}=-i\cos\theta,\\
\\
f_{\hspace{0.1cm}13}^{\prime1}=i\cos\theta,\\
\\
f_{\hspace{0.1cm}10}^{\prime1}=i\sin\theta,\\
\\
f_{\hspace{0.1cm}23}^{\prime2}=-i\cos\theta,\\
\\
f_{\hspace{0.1cm}20}^{\prime2}=-i\sin\theta.
\end{array}
\label{a65}%
\end{equation}
These are quite different from those of the direct product algebra.

The Killing-Cartan bilinear form associated to the group is given by
\begin{equation}
g_{ab}=f_{\hspace{0.1cm}ad}^{c}f_{\hspace{0.1cm}bc}^{d}%
\end{equation}
and its determinant is equal to zero. This characterizes the Glashow algebra
as a non-semisimple one. We have studied other invariant metrics
(Aldrovandi, Barbosa, Crispino and Pereira, 1999), but they all have null determinant.

\section{Representations of the Glashow algebra}

We now present two matrix representations of the Glashow algebra. The first
one to be considered is the adjoint representation which is constructed using
the structure constants $\left(  \ref{a65}\right)  .$ In that case, the
generators are:%
\begin{equation}
J_{1}^{\prime}=\left[
\begin{array}
[c]{cccc}%
0 & 0 & i\cos\theta &  i\sin\theta\\
0 & 0 & 0 & 0\\
0 & -i\cos\theta & 0 & 0\\
0 & -i\sin\theta & 0 & 0
\end{array}
\right] ,\label{b1}%
\end{equation}%
\begin{equation}
J_{2}^{\prime}=\left[
\begin{array}
[c]{cccc}%
0 & 0 & 0 & 0\\
0 & 0 & -i\cos\theta & -i\sin\theta\\
i\cos\theta & 0 & 0 & 0\\
i\sin\theta & 0 & 0 & 0
\end{array}
\right],\label{b2}%
\end{equation}%
\begin{equation}
J_{3}^{\prime}=\left[
\begin{array}
[c]{cccc}%
-i\cos\theta & 0 & 0 & 0\\
0 & i\cos\theta & 0 & 0\\
0 & 0 & 0 & 0\\
0 & 0 & 0 & 0
\end{array}
\right],\label{b3}%
\end{equation}%
\begin{equation}
J_{0}^{\prime}=\left[
\begin{array}
[c]{cccc}%
-i\sin\theta & 0 & 0 & 0\\
0 & i\sin\theta & 0 & 0\\
0 & 0 & 0 & 0\\
0 & 0 & 0 & 0
\end{array}
\right].\label{a59}%
\end{equation}

Another representation can be obtained if we consider the adjoint
representation of the direct product $\left(  \ref{a25}\right)  $-$\left(
\ref{a28}\right)  $ and apply to it the transformations $\left(
\ref{a43}\right)  $-$\left(  \ref{a46}\right)  $ :
\begin{equation}
X_{1}^{\prime}=\frac{1}{\sqrt{2}}\left[
\begin{array}
[c]{cccc}%
0 & 0 & i & 0\\
0 & 0 & -1 & 0\\
-i & 1 & 0 & 0\\
0 & 0 & 0 & 0
\end{array}
\right],\label{b4}%
\end{equation}%
\begin{equation}
X_{2}^{\prime}=\frac{1}{\sqrt{2}}\left[
\begin{array}
[c]{cccc}%
0 & 0 & -i & 0\\
0 & 0 & -1 & 0\\
i & 1 & 0 & 0\\
0 & 0 & 0 & 0
\end{array}
\right],\label{b5}%
\end{equation}%
\begin{equation}
X_{3}^{\prime}=\left[
\begin{array}
[c]{cccc}%
0 & -\cos\theta & 0 & 0\\
\cos\theta & 0 & 0 & 0\\
0 & 0 & 0 & 0\\
0 & 0 & 0 & -i\sqrt{2}\sin\theta
\end{array}
\right],\label{b6}%
\end{equation}%
\begin{equation}
X_{0}^{\prime}=\left[
\begin{array}
[c]{cccc}%
0 & -\sin\theta & 0 & 0\\
\sin\theta & 0 & 0 & 0\\
0 & 0 & 0 & 0\\
0 & 0 & 0 & i\sqrt{2}\cos\theta
\end{array}
\right],\label{a60}%
\end{equation}
These are the transformed matrices of the adjoint representation of the direct
product, which is different from the adjoint representation of the transformed
algebra. The adjoint representation of the transformed algebra does not
coincide with the transformed of the adjoint representation of $SU(2)\otimes
U(1).$ This is the representation which will lead to the Lagrangian as will be
seen is section 5.

There is still another representation for Glashow algebra. It corresponds to
the transformed representation of the fundamental representation of the direct
product:
\begin{equation}
T_{1}^{\prime}=-\frac{i}{\sqrt{2}}\left[
\begin{array}
[c]{ccc}%
0 & 1 & 0\\
0 & 0 & 0\\
0 & 0 & 0
\end{array}
\right] ,
\end{equation}%
\begin{equation}
T_{2}^{\prime}=-\frac{i}{\sqrt{2}}\left[
\begin{array}
[c]{ccc}%
0 & 0 & 0\\
1 & 0 & 0\\
0 & 0 & 0
\end{array}
\right],
\end{equation}%
\begin{equation}
T_{3}^{\prime}=\frac{i}{2}\left[
\begin{array}
[c]{ccc}%
-\cos\theta & 0 & 0\\
1 & \cos\theta & 0\\
0 & 0 & -\sqrt{2}\sin\theta
\end{array}
\right],
\end{equation}%
\begin{equation}
T_{0}^{\prime}=\frac{i}{2}\left[
\begin{array}
[c]{ccc}%
-\sin\theta & 0 & 0\\
1 & \sin\theta & 0\\
0 & 0 & \sqrt{2}\cos\theta
\end{array}
\right].
\end{equation}
This is the single $3\times3$ representation found, and can be considered as
the fundamental representation of the Glashow algebra.

Now a natural question arises: are the $A_{\hspace{0.1cm}\mu}^{\prime}$ gauge
fields for the Glashow algebra, that is, do they belong to the adjoint
representation? To answer this question we must determine the behavior of
$A_{\hspace{0.1cm}\mu}^{\prime}$ under action of the fields $X_{a}^{\prime}.$
We use the transformation $\left(  \ref{a57}\right)  $ of the gauge fields
$A_{\mu}$ and the expressions $\left(  \ref{a32}\right)  ,$ $\left(
\ref{a33}\right)  ,$ $\left(  \ref{a40}\right)  $ of these fields, as well as
$\left(  \ref{a43}\right)  $-$\left(  \ref{a46}\right)  ,$ to obtain%
\begin{equation}%
\begin{array}
[c]{c}%
X_{a}^{\prime}(A_{\hspace{0.1cm}\mu}^{\prime b})=f_{\hspace{0.1cm}ac}^{\prime
b}A_{\hspace{0.1cm}\mu}^{\prime c}.
\end{array}
\end{equation}
The structure constants $\left(  \ref{a65}\right)  $ have also been used.

Thus, the fields $A_{\hspace{0.1cm}\mu}^{\prime}$ are indeed gauge fields. We
may obtain their equations of motion in the usual way for gauge theories, via
the duality prescription applied to the Bianchi identity. Or we may consider
alternatively the usual Lagrangian of gauge theory to describe their dynamics.

The field strength associated to $A_{\hspace{0.1cm}\mu}^{\prime}$ can be
determined using the same argument above $\left(  \ref{a31}\right)  $ to write
the equality
\begin{equation}
F_{\hspace{0.1cm}\mu\nu}^{a}X_{a}=F_{\hspace{0.1cm}\mu\nu}^{\prime a}%
X_{a}^{\prime}.\label{a48}%
\end{equation}
Using the expression of the field strength of $A_{\mu}$ we have
\begin{equation}
g\left\{  \partial_{\mu}A_{\hspace{0.1cm}\nu}^{a}X_{a}-\partial_{\nu}%
A_{\hspace{0.1cm}\mu}^{a}X_{a}+g\left[  A_{\hspace{0.1cm}\mu}^{b}%
X_{b},A_{\hspace{0.1cm}\nu}^{c}X_{c}\right]  \right\}  =F_{\hspace{0.1cm}%
\mu\nu}^{\prime a}X_{a}^{\prime}.
\end{equation}
Now applying $\left(  \ref{a31}\right)  ,$ we obtain
\begin{equation}
g\left\{  \partial_{\mu}A_{\hspace{0.1cm}\nu}^{\prime a}X_{a}^{\prime}%
-\partial_{\nu}A_{\hspace{0.1cm}\mu}^{\prime a}X_{a}^{\prime}+g\left[
A_{\hspace{0.1cm}\mu}^{\prime b}X_{b}^{\prime},A_{\hspace{0.1cm}\nu}^{\prime
c}X_{c}^{\prime}\right]  \right\}  =F_{\hspace{0.1cm}\mu\nu}^{\prime a}%
X_{a}^{\prime},
\end{equation}
which gives
\begin{equation}
F_{\hspace{0.1cm}\mu\nu}^{\prime a}=g\left[ \partial_{\mu}A_{\hspace{0.1cm}\nu
}^{\prime a}-\partial_{\nu}A_{\hspace{0.1cm}\mu}^{\prime a}+gf_{\hspace
{0.1cm}bc}^{\prime a}A_{\hspace{0.1cm}\mu}^{\prime b}A_{\hspace{0.1cm}\nu
}^{\prime c} \right].
\end{equation}
We have proceeded with this punctiliousness because of the above finding,
according to which the adjoint of \textit{GA }is not the transformed of the
adjoint of $SU(2)\otimes U(1).$

With the help of $\left(  \ref{a65}\right)  $ these expressions can be
explicitly written as
\begin{equation}%
\begin{array}
[c]{c}%
F_{\hspace{0.1cm}\mu\nu}^{\prime1}=g\left[\partial_{\mu}A_{\hspace{0.1cm}\nu}%
^{\prime1}-\partial_{\nu}A_{\hspace{0.1cm}\mu}^{\prime1}+ig\cos\theta\left(
A_{\hspace{0.1cm}\mu}^{\prime1}A_{\hspace{0.1cm}\nu}^{\prime3}-A_{\hspace
{0.1cm}\nu}^{\prime1}A_{\hspace{0.1cm}\mu}^{\prime3}\right) \right. \\
\\
+\left. ig\sin\theta\left(  A_{\hspace{0.1cm}\mu}^{\prime1}A_{\hspace{0.1cm}\nu
}^{\prime0}-A_{\hspace{0.1cm}\nu}^{\prime1}A_{\hspace{0.1cm}\mu}^{\prime
0}\right)\right],
\end{array}
\label{a49}%
\end{equation}%
\begin{equation}%
\begin{array}
[c]{c}%
F_{\hspace{0.1cm}\mu\nu}^{\prime2}= g\left[ \partial_{\mu}A_{\hspace{0.1cm}\nu}%
^{\prime2}-\partial_{\nu}A_{\hspace{0.1cm}\mu}^{\prime2}-ig\cos\theta\left(
A_{\hspace{0.1cm}\mu}^{\prime2}A_{\hspace{0.1cm}\nu}^{\prime3}-A_{\hspace
{0.1cm}\nu}^{\prime2}A_{\hspace{0.1cm}\mu}^{\prime3}\right) \right. \\
\\
- \left. ig\sin\theta\left(  A_{\hspace{0.1cm}\mu}^{\prime2}A_{\hspace{0.1cm}\nu
}^{\prime0}-A_{\hspace{0.1cm}\nu}^{\prime2}A_{\hspace{0.1cm}\mu}^{\prime
0}\right)\right],
\end{array}
\label{a50}%
\end{equation}%
\begin{equation}
F_{\hspace{0.1cm}\mu\nu}^{\prime3}= g\left[\partial_{\mu}A_{\hspace{0.1cm}\nu}%
^{\prime3}-\partial_{\nu}A_{\hspace{0.1cm}\mu}^{\prime3}-ig\cos\theta\left(
A_{\hspace{0.1cm}\mu}^{\prime1}A_{\hspace{0.1cm}\nu}^{\prime2}-A_{\hspace
{0.1cm}\nu}^{\prime2}A_{\hspace{0.1cm}\mu}^{\prime1}\right)\right],  \label{a51}%
\end{equation}%
\begin{equation}
F_{\hspace{0.1cm}\mu\nu}^{\prime0}=g\left[ \partial_{\mu}A_{\hspace{0.1cm}\nu}%
^{\prime0}-\partial_{\nu}A_{\hspace{0.1cm}\mu}^{\prime0}-ig\sin\theta\left(
A_{\hspace{0.1cm}\mu}^{\prime1}A_{\hspace{0.1cm}\nu}^{\prime2}-A_{\hspace
{0.1cm}\nu}^{\prime2}A_{\hspace{0.1cm}\mu}^{\prime1}\right)\right].  \label{a52}%
\end{equation}

Once in possession of the field strength, we can now construct the Lagrangian.

\section{Lagrangian}

The Lagrangian for a gauge theory is%
\begin{equation}
L=\frac{1}{8g^2}\int d^{3}x\textnormal{ }tr\left(  F_{\mu\nu}F^{\mu\nu}\right)  ,
\end{equation}
where in the present case $F_{\mu\nu}$ is the field strength in the original
algebra, that is, the direct product algebra. In order to obtain the
expression of the Lagrangian in the Glashow case we use $\left(
\ref{a48}\right)  :$%
\begin{equation}
L=\frac{1}{8g^2}\int d^{3}x\textnormal{ }F_{\hspace{0.1cm}\mu\nu}^{\prime a}%
F_{\hspace{0.1cm}}^{\prime b\mu\nu}tr\left(  X_{a}^{\prime}X_{b}^{\prime
}\right)  ,
\end{equation}
where $X_{a}^{\prime}$ are elements of the transformed representation of the
adjoint algebra of the direct product, $\left(\ref{b4}\right)$ - 
$\left(\ref{a60}\right)$. It is important to notice that here it
is not to be considered the adjoint representation of the \textit{Glashow
algebra }but coherently, the other representation, that is, the transformation
of the adjoint representation of the direct product. If we choose to
calculate the traces in the Lagrangian, the adjoint representation of the
Glashow algebra, we obtain a derivative coupling between the fields $A_{\nu}$
and $Z_{\nu}$ which do not correspond to any vertex of the physical theory.

Thus the role of the second representation is to give the traces for the
Lagrangian. The non null traces are
\begin{equation}%
\begin{array}
[c]{c}%
tr\left(  X_{0}^{\prime}X_{0}^{\prime}\right)  =-2,\\
\\
tr\left(  X_{1}^{\prime}X_{2}^{\prime}\right)  =-2,\\
\\
tr\left(  X_{3}^{\prime}X_{3}^{\prime}\right)  =-2,
\end{array}
\end{equation}
and the resulting Lagrangian is
\begin{equation}
L=-\frac{1}{4g^2}\int d^{3}x\textnormal{ }\left\{  F_{\hspace{0.1cm}\mu\nu}^{\prime
0}F_{\hspace{0.1cm}}^{\prime0\mu\nu}+F_{\hspace{0.1cm}\mu\nu}^{\prime
3}F_{\hspace{0.1cm}}^{\prime3\mu\nu}+2F_{\hspace{0.1cm}\mu\nu}^{\prime
1}F_{\hspace{0.1cm}}^{\prime2\mu\nu}\right\}  .
\end{equation}

Let us substitute $\left(  \ref{a49}\right)  $ -$\left(  \ref{a52}\right)  $
in the last expression. Making the following associations
\begin{equation}%
\begin{array}
[c]{c}%
A_{\hspace{0.1cm}\nu}^{\prime1}\longrightarrow W_{\nu}^{-},\\
\\
A_{\hspace{0.1cm}\nu}^{\prime2}\longrightarrow W_{\nu}^{+},\\
\\
A_{\hspace{0.1cm}\nu}^{\prime3}\longrightarrow Z_{\nu},\\
\\
A_{\hspace{0.1cm}\nu}^{\prime0}\longrightarrow A_{\nu},
\end{array}
\end{equation}
we obtain the electroweak theory Lagrangian without mass:
\begin{equation}%
\begin{array}
[c]{c}%
L=\int d^{3}x\{-\frac{1}{4} {\mathcal{F}}_{\mu\nu}{\mathcal{F}}^{\mu\nu}-\frac
{1}{4}Z_{\mu\nu}Z^{\mu\nu}- \frac{1}{2} W_{\mu\nu}^{+}W^{-\mu\nu}\\
\\
+ig\cos\theta[(W_{\mu}^{-}W_{\nu}^{+}-W_{\nu}^{-}W_{\mu}^{+}%
)\partial^{\mu}Z^{\nu}+(\partial_{\mu}W_{\nu}^{+}-\partial_{\nu}W_{\mu}%
^{+})W^{-\nu}Z^{\mu}\\
\\
-(\partial_{\mu}W_{\nu}^{-}-\partial_{\nu}W_{\mu}^{-})W^{+\nu}Z^{\mu}]\\
\\
+ig\sin\theta[(W_{\mu}^{-}W_{\nu}^{+}-W_{\nu}^{-}W_{\mu}^{+}%
)\partial^{\mu}A^{\nu}+(\partial_{\mu}W_{\nu}^{+}-\partial_{\nu}W_{\mu}%
^{+})W^{-\nu}A^{\mu}\\
\\
-(\partial_{\mu}W_{\nu}^{-}-\partial_{\nu}W_{\mu}^{-})W^{+\nu}A^{\mu}]\\
\\
+g^{2}\cos^{2}\theta[ W_{\mu}^{+}W_{\nu}^{-}Z^{\mu}Z^{\nu}-W_{\mu}%
^{+}W^{-\mu}Z_{\nu}Z^{\nu}]\\
\\
+g^{2}\sin^{2}\theta[ W_{\mu}^{+}W_{\nu}^{-}A^{\mu}A^{\nu}-W_{\mu}%
^{+}W^{-\mu}A_{\nu}A^{\nu}]\\
\\
+g^{2}\sin\theta\cos\theta[ W_{\mu}^{+}W_{\nu}^{-}(Z^{\mu}A^{\nu}%
+A^{\mu}Z^{\nu})-2W_{\mu}^{+}W^{-\mu}A_{\nu}Z^{\nu}]\\
\\
+\frac{1}{2}g^{2}W_{\mu}^{-}W_{\nu}^{+}[W^{-\mu}W^{+\nu}-W^{-\nu}W^{+\mu}]\},
\end{array}
\end{equation}
where
\begin{equation}
\mathcal{F}_{\mu\nu}=\partial_{\mu}A_{\nu}-\partial_{\nu}A_{\mu},
\end{equation}

\begin{equation}
Z_{\mu\nu}=\partial_{\mu}Z_{\nu}-\partial_{\nu}Z_{\mu},
\end{equation}

\begin{equation}
W_{\mu\nu}^{+}=\partial_{\mu}W_{\nu}^{+}-\partial_{\nu}W_{\mu}^{+},
\end{equation}

\begin{equation}
W^{-\mu\nu}=\partial^{\mu}W^{-\nu}-\partial^{\nu}W^{-\mu}.
\end{equation}

\section{Conclusion}

On the bundle of the electroweak theory, we have constructed an algebra 
{\itshape GA} through basis transformations affecting only the 
fundamental, vertical fields. The mixing angle is incorporated in the 
{\itshape GA} structure constants. {\itshape GA} has two representations 
in terms of $4\times4$ matrices. One of them is its adjoint 
representation and the other is the transformed of the adjoint 
representation of the algebra of $SU(2)\otimes U(1)$. It is an 
important point that they do not 
coincide. It is the transformed of the adjoint representation of 
$SU(2)\otimes U(1)$ which provides the 
traces leading to the Lagrangian. The result is just the Lagrangian 
of electroweak interactions without masses.

These massless ``physical'' fields are the gauge fields written in the new basis. 
They are indeed gauge potentials for the Glashow algebra, that is, 
they belong to its adjoint representation. 

Every gauge theory has an underlying bundle which fixes its 
geometrical aspects. The local properties are summed up in the algebra 
of the vector fields tangent to the bundle. {\itshape GA} is that 
algebra for the Glashow model, basic step in the construction of the 
electroweak theory.

\section*{Acknowledgments}

The author thanks Ruben Aldrovandi for enlightening discussions and is 
most grateful to FAPESP (S\~{a}o Paulo, Brazil), for financial 
support.

\section*{References}

\noindent Aldrovandi, R., and Pereira, J.G., (1995). \textit{An Introduction to Geometric Physics}, World
Scientific, Singapore.

\noindent Aldrovandi, R., Barbosa, A.L., Crispino, L.C.B. and Pereira, J.G., (1999). Class. Quantum Grav. \textbf{16} 495.

\noindent Cheng, T.P., and Li, L. F., (1984). \textit{Gauge Theory of 
Elementary Particles}, Oxford University Press.

\noindent Cho, Y.M., (1975). J. Math. Phys. \textbf{16} 2029.

\noindent Greiner, W., and M\"{u}ller, B., (1996). \textit{Gauge Theory of Weak
Interactions }, Springer, New York.

\noindent Mandl, F., and Shaw, G., (1984). \textit{Quantum Field Theory}, John 
Wiley, New York.

\end{document}